\documentclass[twocolumn]{aastex63}
\usepackage{graphicx}	
\usepackage{amsmath}	
\usepackage{amssymb}	
\usepackage{color}
\usepackage{array}
\usepackage{lineno}
\usepackage{bookmark}
\usepackage{subfigure}
\usepackage{booktabs}
\usepackage{float}
\usepackage{threeparttable}

\begin{document}

\title{Anisotropy of core-collapse supernovae effected by AGN disks}

\correspondingauthor{Tong Liu}
\email{tongliu@xmu.edu.cn}

\author{Jiao-Zhen She}
\affiliation{Department of Astronomy, Xiamen University, Xiamen, Fujian 361005, China}

\author[0000-0001-8678-6291]{Tong Liu}
\affiliation{Department of Astronomy, Xiamen University, Xiamen, Fujian 361005, China}

\author[0000-0002-4448-0849]{Bao-Quan Huang}
\affiliation{Department of Astronomy, Xiamen University, Xiamen, Fujian 361005, China}

\author[0000-0002-9130-2586]{Yun-Feng Wei}
\affiliation{Institute of Fundamental Physics and Quantum Technology, Ningbo University, Ningbo, Zhejiang 315211, China}
\affiliation{School of Physical Science and Technology, Ningbo University, Ningbo, Zhejiang 315211, China}

\author[0000-0002-5323-2302]{Fu-Lin Li}
\affiliation{Purple Mountain Observatory, Chinese Academy of Sciences, Nanjing 210023, China}
\affiliation{University of Science and Technology of China, Hefei 230026, China}

\author[0000-0001-9648-7295]{Jin-Jun Geng}
\affiliation{Purple Mountain Observatory, Chinese Academy of Sciences, Nanjing 210023, China}

\author[0000-0002-6299-1263]{Xue-Feng Wu}
\affiliation{Purple Mountain Observatory, Chinese Academy of Sciences, Nanjing 210023, China}
\affiliation{University of Science and Technology of China, Hefei 230026, China}

\begin{abstract}
Active galactic nucleus (AGN) disk provide dense environments to influence on the star formation, evolution, and migration. In AGN disks, pressure gradients and migration accelerations could create the anisotropy of the core-collapse supernovae (CCSNe) when the massive stars explode at the end of their lives. In this study, we construct the equilibrium equations by considering the above two factors and then compute the light curves for three types of progenitor stars at different locations of AGN disks for the different supermassive black hole (SMBH) masses, accretion efficiencies, explosion energies, and masses of ejecta. The results show that the migration acceleration has more significant effects on the anisotropic explosions than the pressure gradients of the AGN disks. The anisotropic luminosities are pronounced at large radii, and massive SMBHs would suppress the anisotropy and reduce the total luminosity.
\end{abstract}


\section{Introduction} \label{sec:intro}

Active galactic nuclei (AGNs) are among the most luminous objects in the universe, powered by the accretion of supermassive black holes (SMBHs). This process releases gravitational energy as mass falls onto the black hole (BH) via an accretion disk \citep{1969Natur.223..690L}, typically described as geometrically thin and optically thick with $\alpha$-viscosity \citep{1973A&A....24..337S}. AGN disks host extreme astrophysical environments which are much hotter and denser than most components of the interstellar medium. These environments are conducive to stellar formation, capture and evolution.

\begin{figure*}[!t]
		\centering
		\includegraphics[width=1.0\linewidth]{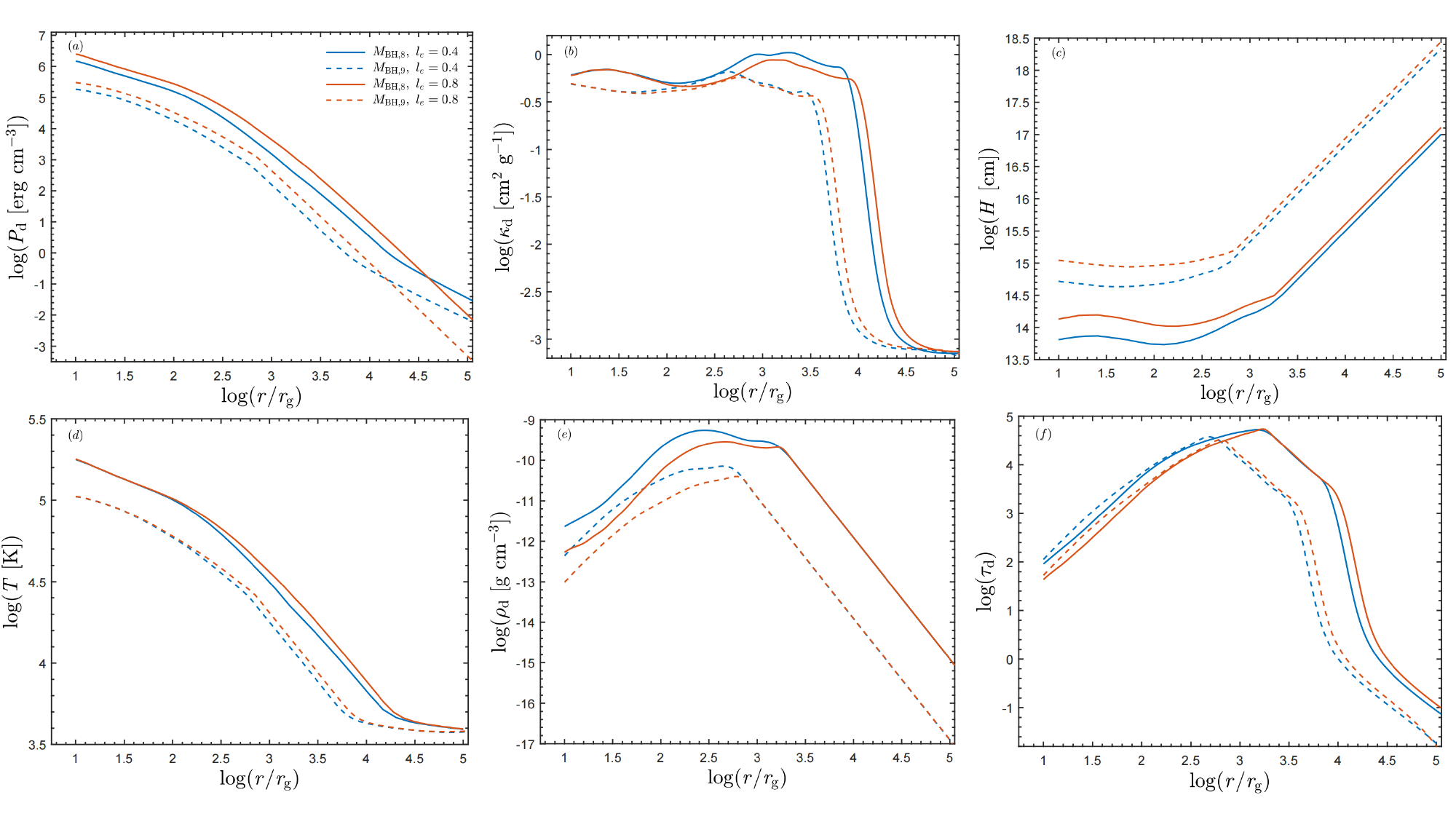}
\caption{Profiles of (a) pressure $P_{\rm{d}}$, (b) opacity $\kappa_{\rm{d}}$, (c) thickness $H$, (d) temperature $T$, (e) density $\rho_{\rm{d}}$, and (f) optical depth $\tau_{\rm{d}}$ of AGN disks with $\alpha=0.1$ and $\varepsilon=0.1$. The solid and dashed lines correspond to $M_{\rm{BH}}=10^8M_{\odot}$ and $10^9M_{\odot}$ ($M_{\rm{BH},8}$ and $M_{\rm{BH},9}$), respectively. The blue and orange lines represent $l_e=0.4$ and $0.8$, respectively.}
\label{fig:1}
\end{figure*}

Gravitationally unstable regions in the disk may trigger vigorous star formation \citep[e.g.,][]{1978AcA....28...91P,2003MNRAS.341..501S,2020MNRAS.493.3732D}, producing accretion-modified stars \citep[e.g.,][]{2021ApJ...911L..14W,2021ApJ...916L..17W} and even long-lived ``immortal'' stars \citep[e.g.,][]{2022ApJ...929..133J}. Interactions between stars and the disk facilitate their capture and embedding \citep[e.g.,][]{1993ApJ...409..592A}, enabling mass growth via gas accretion \citep[e.g.,][]{2021ApJ...910...94C,2020MNRAS.498.3452D} and migration induced by the disk torques \citep[e.g.,][]{1983ApJ...273...99O,1991MNRAS.250..505S}. The high stellar density and gas-rich environment also promote binary formation through tidal interactions \citep[e.g.,][]{1999ApJ...521..502C} or gas-mediated capture \citep[e.g.,][]{2002Natur.420..643G,2020ApJ...898...25T}. These binaries may evolve into stellar mergers \citep[e.g.,][]{2011ApJ...726...28B,2021ApJ...914L..19Z} or gravitational wave-driven compact-object coalescence \citep[e.g.,][]{2012MNRAS.425..460M}. Stellar Evolution in AGN Disks were investigated by \citet[e.g.,][]{2025ApJ...981...16F} and \citet[e.g.,][]{2025ApJ...979..245D}. A significant fraction of these stars can accrete material and undergo core-collapse supernovae (CCSNe) before being accreted by the central SMBH \citep[e.g.,][]{1999A&A...344..433C,2021ApJ...910...94C}.

By analogy with planetary migration, massive stars embedded in AGN disks may also experience inward migration due to disk torques, potentially accelerating under certain conditions. A low-mass planet in a viscous protoplanetary disk undergoes classical Type I migration, typically resulting in inward motion \citep[e.g.,][]{1984ApJ...285..818P,2003ApJ...586..540D}. Giant planets in the Jupiter-mass range carve deep gaps around their orbits and experience Type II migration \citep[e.g.,][]{1986ApJ...309..846L}, while subgiant planets opening partial gaps may undergo Type III migration in massive disks \citep[e.g.,][]{2003ApJ...588..494M,2024ApJ...971..130L}. Similarly, migrating stellar progenitors may experience acceleration. When combined with AGN disk pressure and the star's radiation pressure, this acceleration could generate a radial pressure gradient within the progenitor. Such a gradient might ultimately drive anisotropic stellar winds. However, this process remains unexplored in current supernova (SN) models.

The explosive fate of these stars modifies the AGN disk and produces unique observational signatures. In typical environments, SN light curves (LCs) are well-characterized, with classifications including Type Ia, Ib/c, IIP, IIn, IIb, and IIL \citep[e.g.,][]{1997ARA&A..35..309F}. Type Ia SNe arise from thermonuclear white dwarf explosions \citep[e.g.,][]{1960ApJ...132..565H} or mergers, while the remaining types originate from CCSNe, where LCs are primarily powered by the diffusion of radioactive energy from $^{56}\rm{Ni}$ and $^{56}\rm{Co}$ into homologously expanding ejecta \citep[e.g.,][]{2012ApJ...746..121C}. Analytical models for these processes have been extensively developed \citep[e.g.,][]{1980ApJ...237..541A,1982ApJ...253..785A,1996snih.book.....A}.

However, SN characteristics in AGN disks are altered by the dense gaseous environment. Early study by \citet{1995MNRAS.276..597R} simulated 2D SN explosions within 1 pc of the SMBH, while \citet{2020MNRAS.498.3452D} extended this work with 3D hydrodynamical simulations focusing on disk disruption. Recently, \citet{2021MNRAS.507..156G} developed an analytical model to track the evolution of shock waves propagating through the disk until breakout, and calculated key observables such as peak luminosity, bolometric LCs, and breakout times. \citet{2024ApJ...967...67H} studied the shock propagation and diffuse with energy injections in AGN disks. Moreover, \citet{2023ApJ...950..161L} investigated LCs of CCSNe in AGN disks by using an ejecta-disk interaction model incorporating AGN disk pressure. These studies predominantly assume isotropic explosions with symmetric shockwaves \citep{1982ApJ...258..790C,1999agnf.book.....K}, neglecting the roles of pressure gradients and migration accelerations.

In this study, we investigate the these impact on CCSNe within AGN disks and analyze the resulting LCs. In Section \ref{sec:model}, we describe the AGN disk model and introduce an anisotropic stellar wind dynamics model for migrating progenitors. In Section \ref{sec:LCS}, we compute LCs of CCSNe, incorporating ejecta-wind-disk interactions and nuclear energy contributions. Conclusions and discussion are made in Section 4.

\section{Dynamics}\label{sec:model}

\subsection{AGN disks}

\begin{figure}[!t]
    \centering
    \includegraphics[width=0.3\textwidth]{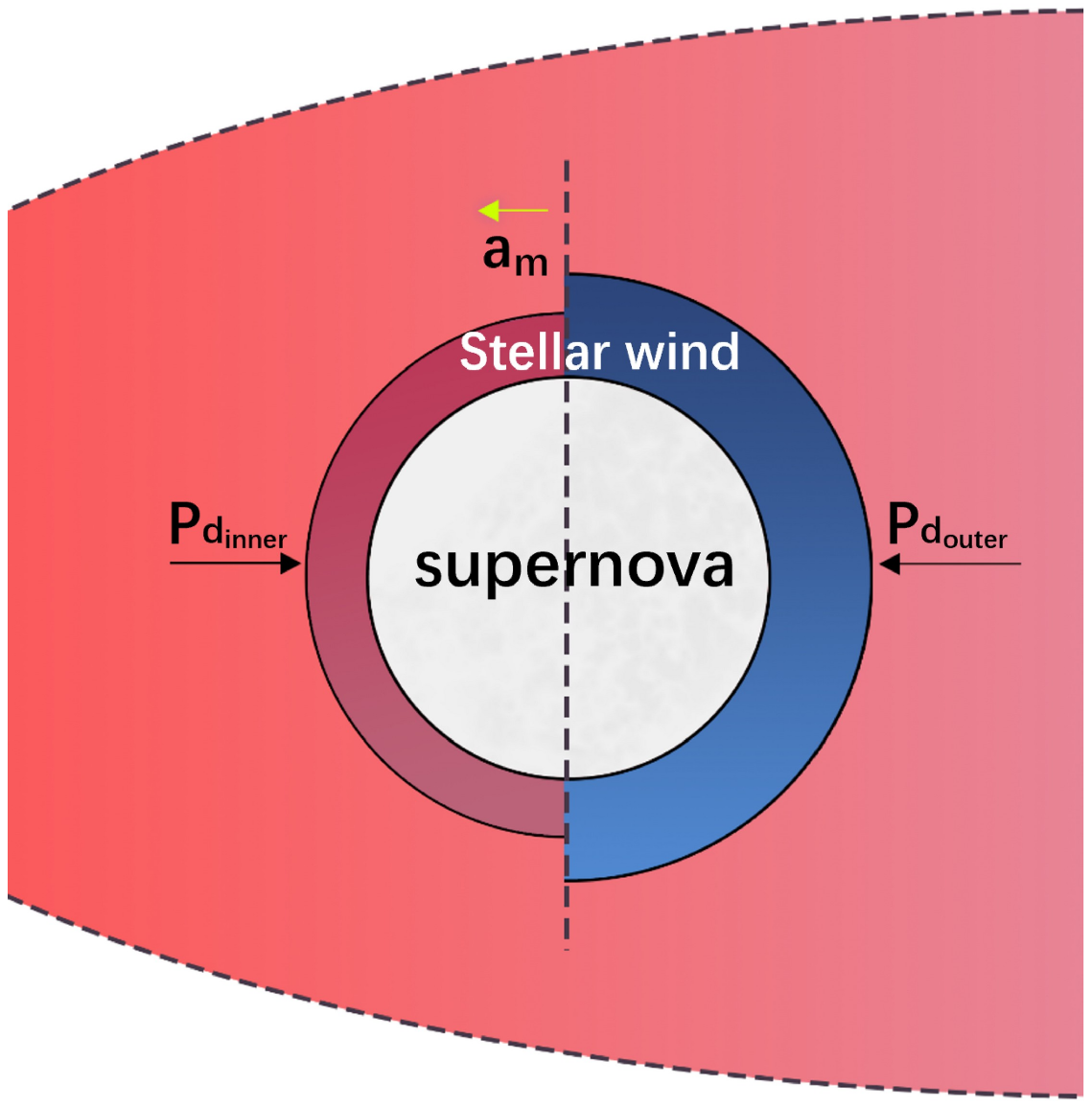}
    \caption{An asymmetric stellar wind under the influence of pressure difference of the AGN disk ($P_{\rm{d_{inner}}}$ and $P_{\rm{d_{outer}}}$) and acceleration of migration ($\rm{a_{\rm{m}}}$).}
    \label{fig:2}
\end{figure}

In this study, we adopt the AGN disk model proposed by \citet{2003MNRAS.341..501S}. Unlike the conventional geometrically thin disk, this model incorporate an additional energy source that powers the outer disk, effectively overcoming self-gravity instability at radius $r \gtrsim 10^{3} ~r_{\mathrm{g}}$, with $r_{\mathrm{g}}= G M_{\rm{BH}} / c^{2}$ is the gravitational radius, $G$ is the gravitational constant, $M_{\rm{BH}}$ denotes the mass of the SMBH, and $c$ is the speed of light.

We assume the outer and inner disk boundaries as $r_{\max}=2 \times 10^{5}~ r_{\mathrm{g}}$ and $r_{\min}= r_{\rm{g}} / 2 \epsilon$, respectively, where $\epsilon = 0.1$ indicates the accretion radiative efficiency. The detailed description of the disk model can be found in \citet{2003MNRAS.341..501S}. Our subsequent analysis relies on the numerical results depicted in Figure~\ref{fig:1}, which contain central SMBH masses of $10^8$ and $10^9~M_{\odot}$, with a viscosity parameter $\alpha = 0.1$. We define $l_{\mathrm{e}} \equiv L_{0} / L_{\mathrm{E}}$ as a measure of the SMBH accretion efficiency, where $L_{\mathrm{E}}$ is the Eddington luminosity, then the accreting rate can be calculated as $\dot{M}_{\rm{BH}}=l_{\mathrm{e}} L_{\mathrm{E}}/ (\epsilon c^2)$. We adopt two SMBH accreting efficiency $l_{\rm{e}}=0.4$ and $0.8$ to investigate the influence of accretion rate on anisotropic explosion.

\subsection{Anisotropic stellar winds}

Massive stars located in the outer regions of AGN disks are highly likely to undergo CCSNe. The environment around these stars significantly differs from the typical interstellar medium in terms of density and temperature. In such environments, more massive stars generally exhibit stronger stellar winds, which can modify the light curve profiles of subsequent CCSNe by increasing the photon diffusion timescale, resulting in smoother LCs. The large diameter of the progenitor star causes the AGN disk to exert different pressures on its two sides, additionally, CCSNe within the disk generally experience migration acceleration, which may generate anisotropic stellar winds. To model this, we derive an equilibrium equation that accounts for the migration acceleration of the progenitor, radiation pressure, and the pressure difference of the AGN disk. Considering the AGN disk pressure is almost not varied in the vertical direction, our analysis focuses only on its effects in radial direction. The simplified diagram is shown in Figure \ref{fig:2}.

\begin{table}[t!]
\begin{footnotesize}
\caption{Parameters of the progenitor}
\label{table1}
\renewcommand\arraystretch{1.2}    
\begin{center}
\setlength{\tabcolsep}{0.5mm}    
\begin{tabular}{@{\extracolsep{\fill}} cccc}
\hline
 Type of progenitor & Progenitor radius ($R_{\odot}$) & Temperature (K)   \\\hline
              RSG   &                          3000   &   $4000$    \\
              BSG   &                           200   &   $3\times10^{4}$  \\
              WR   &                           10    &   $10^{5}$    \\\hline
\end{tabular}
\end{center}
\end{footnotesize}
\end{table}

In general, massive stars can create wind-blown bubbles which drive the expansion of an outer shell under energy conservation \citep{1977ApJ...218..377W}. Recently, numerical simulations and theoretical analyses indicate that adiabatic thermal pressure within the stellar wind bubble is rapidly dissipated in the dense environment, and the bubble expansion becomes almost entirely momentum driven \citep{2021ApJ...914...90L,2023Galax..11...78D}. In the complex environment of AGN disks, however, the processes governing the formation and evolution of stellar wind bubbles remain highly uncertain. The earliest expansion should proceed in an adiabatic and energy-conserving phase, particularly in the disk midplane where the gas is extremely dense and optically thick, which suppresses radiative losses from the bubble. As the cavity grows and radiative cooling strengthens, wind-blown bubbles transitions into a momentum-driven model. A rigorous treatment must follow the evolution from optically thick to optically thin conditions; balance radii obtained by adopting either limit in isolation inevitably carry substantial uncertainties. Typically, energy conserving stellar winds carve out larger bubbles than their momentum driven counterparts. Ignoring the detailed formation history of the bubble, we use momentum-driven model to calculate conservative lower limit of stellar wind balance radii. In addition, given that multiple scattering during photon propagation can enhance the stellar wind momentum by only a factor of $2\sim3$ \citep{1985ApJ...288..679A}, we assume that the photon momentum flow is entirely transferred to the stellar wind material via single scattering. This assumption serves as a simple approximation for momentum transfer in the AGN disk environment \citep{2023ApJ...950..161L}. The wind luminosity is expressed as $L_{s} = \dot{m}_{s} v_{\infty} c$, where $\dot{m}_{s}$ represents the mass loss rate of the progenitor and $v_{\infty}$ is the terminal velocity of the wind. We adopt a spherical configuration to simplify the actual complex AGN disk environment. The momentum flow transferred outward per second can then be calculated as:
\begin{equation}
    \frac{\Delta P}{\Delta t}=\frac{L_{s}}{c}=\frac{4 \pi r_{\rm{p}}^2 \sigma_{\rm{SB}} T_{\mathrm{s,eff}}^4}{c},
	\label{eq:6}
\end{equation}
where $\sigma_{\mathrm{SB}}$ is the Stefan-Boltzmann constant, $r_{\rm{p}}$ and $T_{\mathrm{s,eff}}$ is the radii and typical effective temperatures of the progenitor, respectively. Thus radiation pressure $P_{\rm{f}}$ can be written as
\begin{equation}
P_{\rm{f}}=\frac{L_s}{4\pi r_{\rm{p}}^2 c}.
\label{eq:7}
\end{equation}
As shown in Figure 2, we establish an inertial reference frame centered on the SMBH and introduce a non-inertial frame centered on the progenitor undergoing accelerated migration. Considering progenitors to remain effectively quasi-static in AGN disk \citep{2023ApJ...950..161L}, the governing equations for stellar wind dynamics on both sides of the progenitor star are as follows:
\begin{align}\label{eq_rb}
-P_{\rm{f}}~4\pi r_{\rm{p}}^2+P_{\rm{d_\text{inner}}}~4\pi r_{\rm{b}_{\rm{inner}}}^2 + (-m a_{\rm{m}}) = 0, \nonumber\\
P_{\rm{f}}~4\pi r_{\rm{p}}^2-P_{\rm{d_\text{outer}}}~4\pi r_{\rm{b}_{\rm{outer}}}^2 + (-m a_{\rm{m}}) = 0,
\end{align}
where $P_{\rm{d_{inner}}}$ and $P_{\rm{d_{outer}}}$ are the AGN disk pressures on both sides, respectively. $r_{\rm{b}_{\rm{inner}}}$ and $r_{\rm{b}_{\rm{outer}}}$ denote the balance radii of star wind on both sides of progenitor, respectively, while $a_{\rm{m}}$ represents the migration acceleration. The whole mass $m=m_{\rm{p}}+4\pi/3 \rho_{\rm{w}}(r_{\rm{b}}^3-r_{\rm{p}}^3)$, with $m_p$ denote mass of progenitor and $\rho_{\rm w}$ is stellar wind density. We adopt typical values for the effective temperature $T_{\mathrm{s,eff}}$ and progenitor radius $r_{\mathrm{p}}$ (shown in Table \ref{table1}) to calculate balance radii $r_{\rm{b}_{\rm{inner}}}$ and $r_{\rm{b}_{\rm{outer}}}$. Notably, the disk pressures of the outermost stellar wind on both sides, i.e., $P_{\rm{d_{inner}}}$ and $P_{\rm{d_{outer}}}$ depend on the location of balance radii in AGN disks we adopted \citep{2003MNRAS.341..501S}. We solve for the balance radii vary with the progenitor's location in the AGN disk, and the results are presented in Figure \ref{fig:3}.

In this study, we adopt three representative progenitor stars with different diameters: red supergiants (RSGs), blue supergiants (BSGs) and Wolf-Rayet stars (WRs). We assume that each progenitor type has a mass of $m_{\rm{p}}=40~M_{\odot}$. The investigation of CCSNe is conducted at typical disk radii of $10^{3}$, $10^{4}$, and $10^{5}~r_{\rm{g}}$ for BSGs and WRs. For RSGs, the corresponding radii are $10^{3.5}$, $10^{4}$, and $10^{5}~r_{\rm{g}}$.

\section{LCs}\label{sec:LCS}

In this work, we extend existing models to incorporate a more detailed characterization of anisotropic stellar winds in the progenitor's environment. We account for the differential luminosity contributions from the stellar winds on either side of the progenitor, which has not been considered in past studies. By modeling the evolution of CCSNe LCs as a function of these anisotropic wind profiles, we provide a more comprehensive understanding of the energy transfer and radiation output in CCSNe embedded within an AGN disk. Theoretical studies of SN LCs have been extensively conducted by \citet{1980ApJ...237..541A, 1982ApJ...253..785A}, who demonstrated that their luminosity primarily arises from the radioactive decay of $^{56} \mathrm{Ni}$ to $^{56} \mathrm{Co}$ and subsequently to $^{56} \mathrm{Fe}$. However, in cases where an SN is embedded within a circumstellar medium (CSM), the interaction between SN ejecta and CSM must be considered as an additional contribution to the observed LCs. \citet{1982ApJ...258..790C} and \citet{1994ApJ...420..268C} formulated analytical solutions describing the formation of a forward shock (FS) and a reverse shock (RS) as the supersonic SN ejecta collide with the CSM. This interaction facilitates the conversion of kinetic energy into radiation, thereby serving as an additional energy source influencing the SN LCs. \citet{2023ApJ...950..161L} calculated the LCs of CCSNe within AGN disks in the isotropic case. Here, we extend this framework by incorporating anisotropic stellar winds profiles. Since the thickness of each side remaining nearly constant in our calculation, therefore we use two isotropic bursts to separately quantify the difference in luminosity caused by these anisotropic stellar winds.

\subsection{Anisotropic ejecta-wind-disk interaction}

When an SN explodes in AGN disk, the supersonic ejecta propagates sequentially through two distinct media. First, the ejecta collides with the stellar wind, forming an FS and RS. Second, as the ejecta sweeps up the stellar wind shell, the mixture of SN ejecta and stellar wind material interacts with the AGN disk material, forming a second FS/RS pair. Finally, the shockwave penetrates the photospheric region and become detectable SN LCs.

The SN ejecta density profile is well described by a broken power-law model with two distinct indices \citep{1982ApJ...258..790C,2023ApJ...950..161L}, which are
\begin{align}
    \rho_{\text{outer}}&=g^n t^{n-3} r^{-n} , \label{eq:9} \\
    \rho_{\text{inner}}&=g^m t^{m-3} r^{-m},  \label{eq:10}
\end{align}
where $n$ and $m$ denote the density indices of the outer and inner ejecta regions, respectively. $g^n$ and $g^m$ are the scaling parameters, which are
\begin{align}
    &g^n=\frac{1}{4 \pi(n-m)} \frac{\left[2(5-m)(n-5) E_{\mathrm{SN}}\right]^{(n-3) / 2}}{\left[(3-m)(n-3) M_{\mathrm{ej}}\right]^{(n-5) / 2}}, \label{eq:11} \\
    &g^m=\frac{1}{4 \pi(n-m)} \frac{\left[(3-m)(n-3) M_{\mathrm{ej}}\right]^{(5-m) / 2}}{\left[2(5-m)(n-5) E_{\mathrm{SN}}\right]^{(3-m) / 2}},
    \label{eq:12}
\end{align}
where $E_{\mathrm{SN}}$ refers to the total energy released by the SN explosion, and $M_{\mathrm{ej}}$ is the ejecta mass. Based on the principles of mass and energy conservation, the velocity $v_{\mathrm{SN}}$ at the break surface of the SN ejecta can be uniquely derived from the parameters $E_{\mathrm{SN}}$, $M_{\mathrm{ej}}$, $m$, and $n$, which can be written as \citep{2012ApJ...746..121C}
\begin{equation}
    v_{\mathrm{SN}}=\left[\frac{(10-2m)(n-5)E_{\mathrm{SN}}}{(3-m)(n-3)M_{\mathrm{ej}}}\right]^{1/2}/x_0,
    \label{eq:13}
\end{equation}
where $x_{0}=r_{\mathrm{core}}(t)/r_{\mathrm{SN}}(t)$ is the dimensionless radius of the break in the SN ejecta density profile from the inner flat component (controlled by $m$) to the outer, steeper component (controlled by $n$), which is at radius $r_{\mathrm{core}}(t)$, $r_{\mathrm{SN}}(t)$ is the radius of the supernova ejection. According to hydrodynamical simulations in  \citet{1994ApJ...420..268C}, \citet{1990ApJ...360..242S}, and \citet{2023ApJ...950..161L}, the ejecta profile can be defined with the parameters $n=7$ and $m=0$; and we set $x_0=0.1$ for RSGs, $0.3$ for BSGs, and $0.9$ for WRs as representative values, where $x_0$ affects the ejecta's peripheral velocity and the energy input timescale for ejecta-wind interaction, but this interaction contributes only a small portion to the final CCSNe LCs.

The profile of stellar wind can be described as $\rho_{\mathrm{w}}=q_{\mathrm{w}} r^{-s_{\mathrm{w}}}$, where $q_{\mathrm{w}}$ is a scaling constant, and $s_{\mathrm{w}}$ is the power-law index of stellar winds. The range of $s$ is $0 \leqslant s_{\mathrm{w}}\leqslant 2$, for $s_{\mathrm{w}}=2$ indicates a steady-wind model and $s_{\mathrm{w}}=0$ indicate the presence of ``bubbles'' or shells formed by strong stellar winds, here we adopt $s_{\mathrm{w}}=0$. In general, if the mass-loss rate $\dot{m}_{\rm w}$ and wind terminal velocity $v_{\rm{w}}$ are known, then we have $q_{\mathrm{w}}=\dot{m}_{\rm w} /\left(4 \pi v_{\mathrm{w}}\right)$. Here we adopt the cell density at outermost radius as the stellar wind density, which take $\rho_{\mathrm{w}}=10^{-11}\rm{g~cm^{-3}}$ based on the simulation of massive star evolution \citep{2007PhR...442..269W}.

The self-similar solutions for the radii of the FS and the RS was found by \citet{1982ApJ...258..790C}, as a function of time:
\begin{align}
    r_{\mathrm{FS},i}(t)=r_{i}+\beta_{\mathrm{FS},i} \left(\frac{A_{{i}} g^n_{{i}}}{q_{{i}}}\right)^{\frac{1}{n-s_{{i}}}} t^{\frac{(n-3)}{\left(n-s_i\right)}},\\
    r_{\mathrm{RS},i}(t)=r_{i}+\beta_{\mathrm{RS},i} \left(\frac{A_{{i}} g^n_{{i}}}{q_{{i}}}\right)^{\frac{1}{n-s_{{i}}}} t^{\frac{(n-3)}{\left(n-s_i\right)}},
    \label{eq:17}
\end{align}
where $\beta_{\mathrm{FS},i}$, $\beta_{\mathrm{RS},i}$, and $A_i$ are constants that depend on the values of $n$ and $s_i$. According to \citet{1982ApJ...258..790C} and \citet{2023ApJ...950..161L}, we set $s_1=s_2=0$ to represent a uniform wind shell and disk material, which gives us $\beta_{\mathrm{FS},i}=1.181$, $\beta_{\mathrm{RS},i}=0.935$, and $A_i=1.2$. Here $i=1$ stands for ejecta-wind interaction while $i=2$ stands for mixture of ejecta and wind interacts with disk material. Moreover, the first interaction occurs at the radius $r_{1}=r_{\mathrm{p}}$, where the SN ejecta first will collide with the stellar wind shell. The second interaction occur at the balance radius $r_{2}=r_{\mathrm{b}}$, where the shock wave sweeps up the wind shell. At this moment, SN ejecta has swept up the whole stellar wind material, the mass of the ejecta should be updated to account for stellar wind mass and $E_{\mathrm{SN}}$ should also be recalculated accordingly, which are
\begin{align}
    M_{\mathrm{ej},2}&=M_{\mathrm{ej},1}+M_{\mathrm{w}}, \label{eq:20}\\
    E_{\mathrm{SN},2}&=E_{\mathrm{SN},1}-E_{\mathrm{rad},1}, \label{eq:21}
\end{align}
where $M_{\mathrm{ej},1}$ and $M_{\mathrm{ej},2}$ represent the ejecta mass during the first and second interaction, respectively; $M_{\mathrm{w}}$ is the mass of stellar wind; $E_{\mathrm{SN,1}}$ and $E_{\mathrm{SN,2}}$ denote the total energy of the first and second interaction, respectively; and $E_{\mathrm{rad},1}$ corresponds to the radiative energy loss during the first interaction. It is notably that the ejecta with $E_{\mathrm{SN,1}}$ is the total energy of SN ejecta which comprises both kinetic and thermal energy, as the ejecta expands, its thermal energy is converted into kinetic energy. In our model, the ejecta and wind shell are treated as a single whole system, the only source of dissipation arises from radiation emitted by this entire system. Furthermore, we neglected any complex energy exchange processes between the ejecta and wind shell during FS or RS events.

In the process of shock wave interaction with two media, the action durations of FS ($t_{\mathrm{FS},i}$) and RS ($t_{\mathrm{RS},i}$) can be written as \citep[e.g.,][]{2012ApJ...746..121C,2023ApJ...950..161L}
\begin{eqnarray}
    &&t_{\mathrm{FS},i} = \nonumber \\
    &&\left[\frac{\left(3-s_i\right) q_i^{(3-n) /\left(n-s_i\right)}\left(A_i g^n_{i}\right)^{\left(s_i-3\right) /\left(n-s_i\right)}}{4 \pi \beta_{\mathrm{FS}, i}^{3-s_i}}\right]^{\frac{n-s_i}{(n-3)\left(3-s_i\right)}} \nonumber \\
    &&\times M_{i}^{\frac{n-s_i}{(n-3)\left(3-s_i\right)}},\label{eq:25}\\
    &&t_{\mathrm{RS}, i}=\nonumber\\
    &&\left[\frac{v_{\mathrm{SN},i}}{\beta_{\mathrm{RS}, i}\left(A_i g^n_{i} / q_i\right)^{\frac{1}{n-s_i}}}\left(1-\frac{(3-n) M_{\mathrm{ej},i}}{4 \pi v_{\mathrm{SN}, i}^{3-n} g^n_{i}}\right)^{\frac{1}{3-n}}\right]^{\frac{n-s_i}{s_i-3}},\label{eq:26}\nonumber \\
\end{eqnarray}
with $t_{\rm{FS,1}}$ is the time of first FS has swept up all the stellar wind material, thus $M_{1}=M_{\mathrm{w}}$, which is
\begin{eqnarray}
    M_{\rm{w}}&=&\int_{r_{\rm{p}}}^{r_{\rm{b}}} 4 \pi r^2 \rho_{\rm{w}} dr.\label{eq:29}
\end{eqnarray}
While $t_{\rm{FS,2}}$ is the time of second FS has swept up to the point where the photon behind the shock diffuses faster than the shock, generally considering $M_{2}=M_{\mathrm{d,2}}$, with $M_{\mathrm{d,2}}$ representing the optically thick part of the disk shell mass, i.e., radiatively opaque regions in the accretion disk, which is
\begin{eqnarray}
    M_{\rm{d,2}}&=&\int_{r_{\rm{b}}}^{r_{\rm{ph}}} 4 \pi r^2 \rho_{\rm{d}} dr, \label{eq:30}
\end{eqnarray}
where $r_{\rm{ph}}$ represents the photospheric radius of the AGN disk. The Eddington approximation is utilized to calculate this parameter, where $r_{\rm{ph}}$ satisfies
\begin{equation}
    \tau=\int_{r_{\mathrm{ph}}}^{H} \kappa_{\rm{d}} \rho_{\mathrm{d}} d r=\frac{2}{3}.
    \label{eq:28}
\end{equation}
Moreover, the total mass of disk material are
\begin{equation}
     M_{\rm{d}}=\int_{r_{\rm{b}}}^{H} 4 \pi r^2 \rho_{\rm{d}} dr. \label{eq:31}
\end{equation}

The luminosity input of FS and RS from two sequential interactions are \citep[e.g.,][]{2012ApJ...746..121C,2023ApJ...950..161L}
\begin{eqnarray}
    L_{\mathrm{FS}, i}(t)= &&\frac{2 \pi}{\left(n-s_i\right)^3} g^{n^{\frac{5-s_i}{n-s_i}}}_{i} q_i^{\frac{n-5}{n-s_i}}(n-3)^2(n-5) \beta_{\mathrm{FS}, i}^{5-s_i}   \nonumber \\
    &&\times A_i^{\frac{5-s_i}{n-s_i}}\left(t+t_{\mathrm{int}, i}\right)^{\frac{\left(2 n+6 s_i-n s_i-15\right)}{\left(n-s_1\right)}} \theta\left(t_{\mathrm{FS}, i}-t\right),\label{eq:23}\nonumber\\\\
    L_{\mathrm{RS}, i}(t)=&& 2 \pi\left(\frac{A_1 g^n_{i}}{q_i}\right)^{\frac{5-n}{n-s_i}} \beta_{\mathrm{RS}, i}^{5-n} g^n_{i}\left(\frac{n-5}{n-3}\right)\left(\frac{3-s_i}{n-s_i}\right)^3  \nonumber \\
    && \times\left(t+t_{\mathrm{int}, i}\right)^{\frac{\left(2 n+6 s_i-n s_i-15\right)}{\left(n-s_i\right)}} \theta\left(t_{\mathrm{RS}, *, i}-t\right),    \label{eq:24}
\end{eqnarray}
where $\theta(t_{\mathrm{RS},*,i}-t)$ and $\theta(t_{\mathrm{FS},i}-t)$ are the Heaviside step function that accounts for the starting time of the energy input of FS and RS. For $i = 1$, the interaction represents the first encounter between the ejecta and the wind, while for $i = 2$, it corresponds to the second interaction involving the ejecta with the disk. The timing of the first interaction between the ejecta and wind shell, $t_{\rm{int,1}}=r_{1}/v_{\mathrm{SN},1}$. After a time delay $t_{\rm{int},2} \simeq r_{2}/v_{\mathrm{SN},2}$, the second interaction between the ejecta and disk begins. The total shock luminosity input from FS and RS is
\begin{eqnarray}
    L_{\mathrm{inp},i}(t)=L_{\mathrm{FS},i}(t)+L_{\mathrm{RS},i}(t).
\end{eqnarray}
Here we calculate the shock luminosity that emerges from the disk photosphere, which provides an upper limit on the luminosity of SN explosions embedded within the AGN disk.

Assuming that the photosphere of the AGN disk is near the surface and well above the SN explosion location, the bolometric SN LC can be written as
\begin{equation}
    L_i(t)=\frac{1}{t_{\mathrm{diff},i}} \exp \left(-\frac{t}{t_{\mathrm{diff},i}}\right) \int_0^t \exp \left(\frac{t^{\prime}}{t_{\mathrm{diff},i}}\right) L_{\mathrm{inp},i}\left(t^{\prime}\right) dt^{\prime},
\end{equation}
with $t_{\mathrm{diff}, 1}$ is the photon diffusion time in the stellar wind shell and the AGN disk, which is
\begin{equation}
    t_{\mathrm{diff}, 1}=\frac{\kappa_{\mathrm{w}} M_{\mathrm{w}}+\kappa_{\mathrm{d}} M_{\mathrm{d},2}}{\beta c r_{\mathrm{ph}}},
    \label{eq:33}
\end{equation}
where $\beta\sim13.8$ is a constant for variable density distribution \citep{1980ApJ...237..541A}, $\kappa_{\mathrm{w}}$ represents the opacity of stellar wind, and $\kappa_{\mathrm{d}}$ denotes the material opacity at progenitor's location within the AGN disk. The term $t_{\mathrm{diff}, 2}$ corresponds to the photon diffusion time in the AGN disk, which is given by
\begin{equation}
    t_{\mathrm{diff}, 2}=\frac{\kappa_{\mathrm{d}} M_{\mathrm{d,2}}}{\beta c r_{\mathrm{ph}}}.
    \label{eq:34}
\end{equation}
These diffusion timescales are approximations under idealized conditions where energy input is centrally directed, with $r_{\mathrm{ph}} \gg r_{\mathrm{FS},i}$ and $r_{\mathrm{RS},i}$.

\begin{figure*}[t!]
		\centering
		\includegraphics[width=1.0\linewidth]{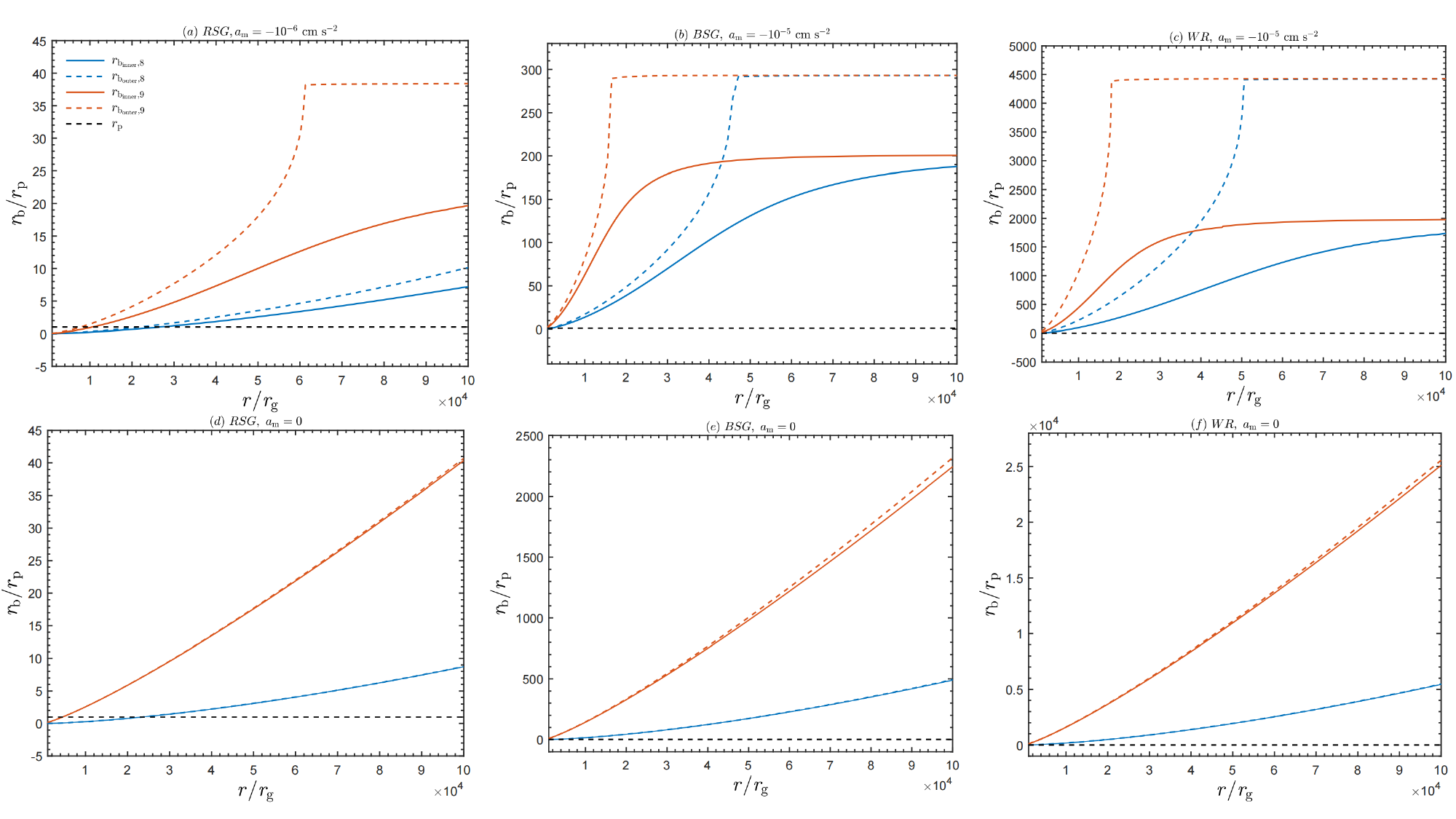}
\caption{The relative thickness of stellar wind for different types of progenitors in the AGN disk with $l_e=0.4$. The top panels (a-c) show balance radius of progenitors with a migration acceleration (RSG: $-10^6~\rm{cm~s^{-2}}$; BSG and WR: $-10^5~\rm{cm~s^{-2}}$), while the bottom panels (d-f) depict cases without acceleration. The solid and dashed lines represent the balance radius on the inner and outer side, respectively. The blue and orange lines represent $M_{\rm{BH},8}$ and $M_{\rm{BH},9}$, respectively, while black dash line denotes progenitors' radius.}
\label{fig:3}
\end{figure*}

\begin{figure*}[tb]
		\centering
		\includegraphics[width=1.4\linewidth]{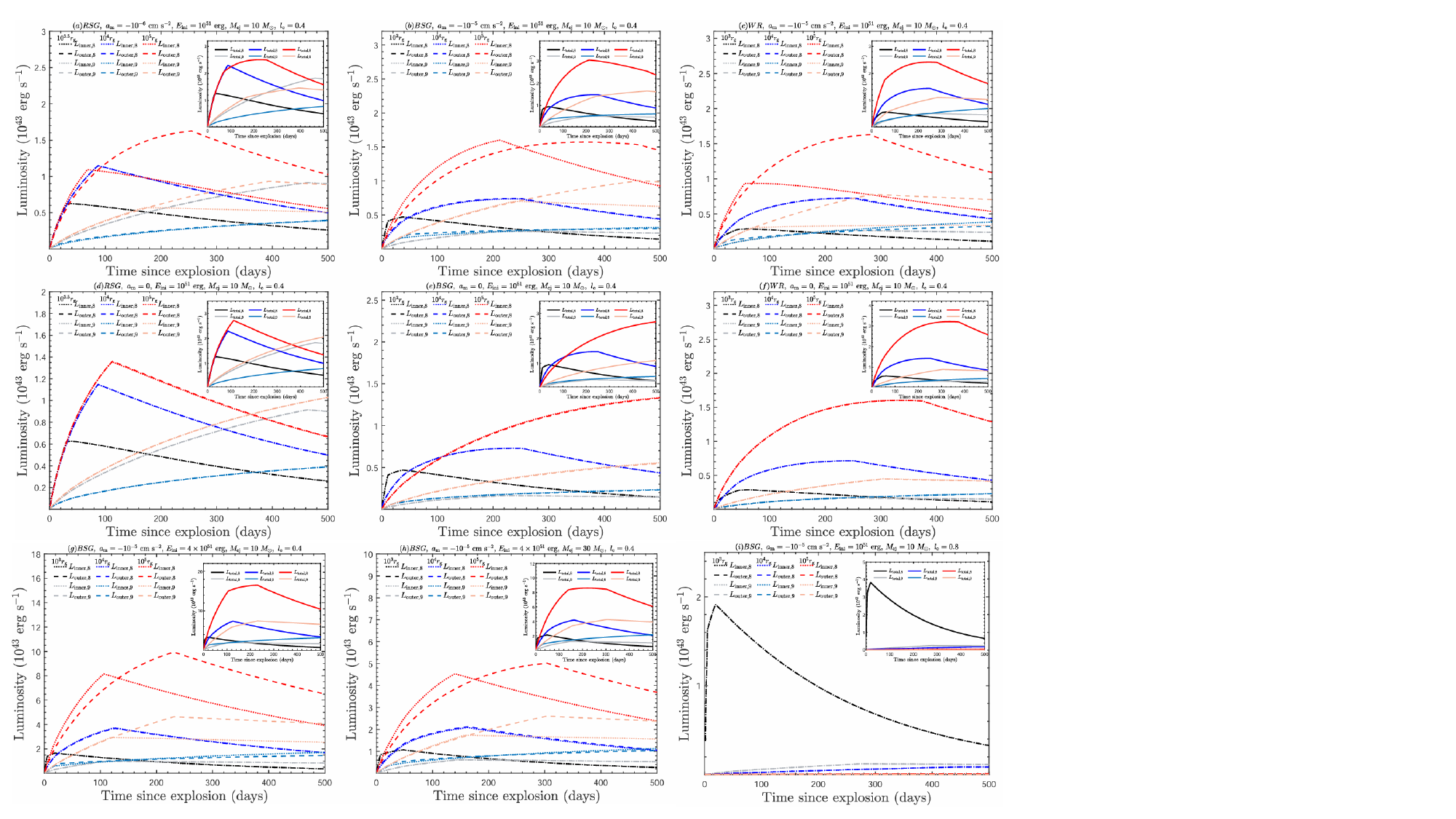}
\caption{The top panels (a-c) show the LCs of CCSNe progenitor with migration acceleration, while the middle panels (d-f) depict cases without acceleration, and the bottom panels (g-i) show the luminosity of BSGs with different $E_{\rm{sn}}$, $M_{\rm{ej}}$, and $l_{\rm{e}}$. The dotted and dashed lines represent the LCs of the inner and outer side, respectively. The dark and light lines represent $M_{\rm{BH},8}$ and $M_{\rm{BH},9}$, respectively. The insets in each figure shows the total luminosities.}
\label{fig:4}
\end{figure*}

Taking into account the anisotropic stellar wind, which consists of contributions from both the inner and outer sides, the total luminosity from shock wave can be expressed as
\begin{equation}
    L_{\rm{t,inp}}(t)=\frac{1}{2}L_{\mathrm{inner},i}(t)+\frac{1}{2}L_{\mathrm{outer},i}(t).
    \label{eq:032}
\end{equation}
where $L_{\mathrm{inner,i}(t)}$ and $L_{\mathrm{outer,i}(t)}$ denote the global luminosities from the inner and outer sides of CCSNe, respectively.

\subsection{Luminosity of~$^{56} \mathrm{Ni}$ decays from ejecta}

We briefly considered nuclear radiation $^{56}$Ni -- $^{56}$Co -- $^{56}$Fe, which follows the decay reaction power \citep[e.g.,][]{2012ApJ...746..121C}
\begin{eqnarray}
    &&L_{\mathrm{nuc}}(t)=\frac{1}{t_{\mathrm{diff,0}}} e^{-\frac{t}{t_{\mathrm{diff,0}}}} \nonumber \\
    &&\int_0^t e^{\frac{t^{\prime}}{t_{\mathrm{diff,0}}}} M_{\mathrm{Ni}}\left[\left(\epsilon_{\mathrm{Ni}}-\epsilon_{\mathrm{Co}}\right) e^{-t^{\prime} / \mathrm{t}_{\mathrm{Ni}}}+\epsilon_{\mathrm{Co}} e^{-t^{\prime} / \mathrm{t}_{\mathrm{Co}}}\right] dt^{\prime}, \nonumber \\
    \label{eq:35}
\end{eqnarray}
where $t_{\mathrm{Ni}}=7.605\times10^5~\mathrm{s}$ and $t_{\mathrm{Co}}=9.822\times10^6\mathrm~{s}$ are the e-folding lifetime of $\mathrm{Ni}$ and $\mathrm{Co}$, respectively. The corresponding energy generation rates are $\epsilon_{\mathrm{Ni}}=3.9\times10^{10} \mathrm{\ erg\ s^{-1}\ g^{-1}}$ and $\epsilon_{\mathrm{Co}}=6.8\times10^9 \mathrm{\ erg\ s^{-1}\ g^{-1}}$. The photon diffusion time $t_{\mathrm{diff,0}}$ across the SN ejecta, stellar wind shell and AGN disk is described by
\begin{equation}
    t_{\mathrm{diff}, 0}=\frac{\kappa_{\mathrm{ej}} M_{\mathrm{ej,1}}+\kappa_{\mathrm{w}} M_{\mathrm{w}}+\kappa_{\mathrm{d}} M_{\mathrm{d}, 2}}{\beta c r_{\mathrm{ph}}},
    \label{eq:36}
\end{equation}
where $\kappa_{\rm{ej}}$ is the opacity of the SN ejecta. The diffusion timescale given here is also an approximation since it assumes that all $^{56}\mathrm{Ni}$ is located in the center of the ejecta. The nuclear-powered luminosity is approximately one order of magnitude lower than the shock-powered one \citep{2023ApJ...950..161L}, which is not dominant term.

Thus, considering the contributions from both the inner and outer stellar wind components around CCSNe, the total luminosity from radioactive decays is given by
\begin{equation}
    L_{\rm{t,nuc}}(t)=\frac{1}{2}L_{\rm{inner,nuc}}(t)+\frac{1}{2}L_{\rm{outer,nuc}}(t),
    \label{eq:32}
\end{equation}
where $L_{\rm{inner,nuc}}(t)$ and $L_{\rm{outer,nuc}}(t)$ represent the global contributions from the inner and outer stellar wind regions to the total radioactive decay luminosity.

\begin{table}[!t]
    \begin{footnotesize}
    \caption{Parameters in CCSNe LCs calculations}
    \label{table:2}
    \renewcommand\arraystretch{1.5}    
    \begin{center}
    \setlength{\tabcolsep}{0.8mm}    
    \begin{tabular}{@{\extracolsep{\fill}} llc}
    \hline
    Parameter  &  Symbol   &  Value  \\\hline
    $\text{SMBH mass}\ ({{M_{\odot}}}) $                    & $M_{\mathrm{BH}}       $ &  8, 9\\
    $\text{Accretion efficiency}     $                    & ${l_\mathrm{e}}$ &  0.4, 0.8\\
    $\text{Mass of progenitors}\ ({{M_{\odot}}})$                    & $M_{\mathrm{p}}$         &  40\\
    $\text{SN explosion energy}\ (10^{51}\mathrm{erg})$                     & $E_{\mathrm{SN}}$        &  1, 4\\
    $\text{SN ejecta mass}\ ({{M_{\odot}}})$                         &  $M_{\mathrm{ej}}$       &  $10,~30$\\
    $\text{$^{56}$Ni mass}\ ({{M_{\odot}}})$                                &  $M_{\mathrm{Ni}}$       &  $1$ \\
    $\text{SN ejecta outer section index} $                                 &  $n$                       &  7\\
    $\text{SN ejecta inner section index}      $                            &  $m$                       &  0\\
    $\text{Stellar wind index}  $                                           &  $s_1$                   &  0\\
    $\text{Disk index}      $                                               &  $s_2$                   &  0\\
    $\text{Stellar wind density scaling parameter}\ (\mathrm{g\ cm^{-3}})$  & $\rho_{\rm{w}}$          &  $10^{-11}$   \\
    $\text{Dimensionless separating radius of RSGs}$                         &  $x_0$                   &  0.1 \\
    $\text{Dimensionless separating radius of BSGs}$                         &  $x_0$                   &  0.3 \\
    $\text{Dimensionless separating radius of WRs}$                         &  $x_0$                   &  0.9 \\
    $\text{Opacity of SN ejecta}\ (\mathrm{cm^2\ g^{-1}})$              &  $\kappa_\mathrm{ej}$    &  0.1\\
    $\text{Opacity of stellar wind}\ (\mathrm{cm^2\ g^{-1}})$           & $\kappa_\mathrm{w}$      &  0.2 \\
    \hline
    \label{table:3}
    \end{tabular}
    \end{center}
    \end{footnotesize}
    \end{table}

By accounting for all energy contributions, the total bolometric LC of the CCSN $L_{\mathrm{total}}(t)$ follows the expression
\begin{equation}
    L_{\mathrm{total}}(t)=L_{\rm{t,inp}}(t)+L_{\mathrm{t,nuc}}(t).
    \label{eq:37}
\end{equation}

\subsection{Results}

We calculate the relative thickness of anisotropic stellar wind for different types of accelerated migrating CCSNe progenitor (RSGs, BSGs, and WRs) vary with location in AGN disk with different central BH mass ($M_{\rm{BH}}=10^8$ and $10^9~M_{\odot}$). These calculations based on the AGN model \citep{2003MNRAS.341..501S} and Equation (\ref{eq_rb}), with result shown in Figure \ref{fig:3}. Additionally, we compute the anisotropic LCs of these progenitors at different radii within the AGN disk, following the method outlined in Section \ref{sec:LCS}. The resulting LCs are presented in Figure \ref{fig:4}. The parameters adopted in our calculations are listed in Table \ref{table:2}.

In Figure \ref{fig:3}, we show the balance radius of anisotropic stellar wind for three types of progenitors with $l_{\rm{e}}=0.4$. The top panels (a-c) show balance radius of progenitors with a migration acceleration (RSGs: $-10^6~\mathrm{cm~s^{-2}}$; BSGs and WRs: $-10^5~\mathrm{cm~s^{-2}}$), while the bottom panels (d-f) show stable progenitors without acceleration. The results show that the three types of progenitors can produce anisotropic stellar winds in both cases. In each case, the outer stellar wind is generally thicker than the inner side and its difference is greater in the outer zone, regardless of whether acceleration is present. This difference is significantly larger in the presence of acceleration case than the non-accelerated case. Moreover, central BH mass of $10^9~M_{\odot}$ is conducive to the production of anisotropic stellar winds at small radii compared to SMBH mass of $10^8~M_{\odot}$. For all progenitors, thicker stellar winds and larger differences between the inner and outer sides are observed at large radii. Specifically, the stellar winds of RSGs is too weak to form within \(\approx 2.5 \times 10^4 ~r_{\mathrm{g}}\) in both cases, consistent with \citet{2023ApJ...950..161L}, whereas BSGs and WRs sustain stellar winds across the explored radii. Notably, in accelerated migrating case, the stellar wind's thickness approaches a saturation as shown in Figure 3 when the pressure contribution becomes negligible compared to accelerated migratory term in Equation (\ref{eq_rb}).

In Figure \ref{fig:4}, we show the LCs of anisotropic explosions for three types of progenitors. The top panels (a-c) show the LCs of progenitors with migration acceleration, while middle panels (d-f) show the LCs of stable progenitors without acceleration, and the bottom panels (g-i) shows the LCs of BSGs for different $E_{\rm{sn}}$, $M_{\rm{ej}}$, and $l_{\rm{e}}$, and the insets in each panels show the total luminosity of inner and outer sides. The results reveal a significant difference between the LCs of the inner and outer side in CCSNe with migration acceleration (a-c). Particularly at $10^5~r_{\rm{g}}$ , the differences of luminosities up to $\sim50\%$, whereas no such difference is observed in CCSNe without migration acceleration (d-f).  Notably, in the location of $r = 10^5~r_{\rm{g}}$, where anisotropic luminosity is pronounced, we can find the duration of outer sides is longer than inner side for both cases. This is because the FS propagate through quickly, and the majority of the energy input comes from the RS.

For RSGs with migration acceleration, no stellar wind is generated at $10^3$ or $10^4~r_{\rm{g}}$, resulting in no LCs differences between the inner and outer sides. However, at $10^5~r_{\rm{g}}$, anisotropic stellar winds form on both sides, leading to a significant luminosity differences. For the total luminosity of inner and outer regions, systems with a central BH of $10^8~M_{\odot}$ exhibit higher values than those with a BH of $10^9~M_{\odot}$. This trend is most pronounced in the outermost regions, where a geometrically thicker disk suppresses radiative efficiency, leading to reduced luminosity, a same behavior also observed in BSGs and WRs.

For BSGs with migration acceleration, anisotropic stellar winds can form on both sides at all positions. However, since the difference in stellar wind thickness between location at $10^3$ and $10^4~r_{\rm{g}}$ is relatively small, the LCs differences between the inner and outer sides is minimal ($\sim2\%$). Moreover, a significant luminosity difference also only appears at $10^5~r_{\rm{g}}$.

For WRs with migration acceleration, anisotropic stellar winds with sufficient differences in thickness can form on both sides at all radii. However, due to the limited scale of WRs, the luminosity difference between the inner and outer sides does not emerge at $10^3$ or $10^4~r_{\rm{g}}$. Similar to RSGs and BSGs, a significant luminosity difference only appears at $10^5~r_{\rm{g}}$.

In contrast to the acceleration cases, non-acceleration RSGs, BSGs, and WRs exhibit no significant differences in LCs between the inner and outer sides, which indicts that the pressure gradients of the AGN disks are negligible roles to effect the anisotropy of CCSNe. The trend of total luminosity is the same as accelerated migration cases, as demonstrated in panels (d-f) for all tested parameters.

In panels (g-i), we can find an increase in the initial energy enhances the total luminosity but does not amplify the luminosity difference between the inner and outer side, as illustrated in Figure 4(g). Conversely, an increase in the ejected mass reduces the total luminosity but also does not enhance the luminosity difference, as shown in Figure 4(h). Moreover, An increase in $l_{\rm{e}}$ leads to an abnormally large disk thickness, which not only reduces the total luminosity but also eliminates the luminosity difference between the two sides, as shown in Figure 4(i).

\section{Conclusions and Discussion}\label{sec:Discussion and conclusions}

In this work, we investigated the anisotropy of CCSNe effected by AGN disks. We calculated how the stellar wind thickness on both sides of the three types of progenitors, i.e., RSGs, BSGs, and WRs, varies with radius by constructing equilibrium equations that account for the migration acceleration of the progenitor and the pressure gradients of the AGN disk. We extend the ejecta-wind-disk model to calculate the anisotropic LCs of those three types of progenitors, which consider several effects, including with or without migration acceleration, locations in AGN disks, central SMBH masses, explosion energies, masses of ejecta, and accretion efficiencies.

In the accelerated migrating CCSNe case, the pressure terms in Equation (\ref{eq_rb}) decreases with the increasing radius of the AGN disk, causing the accelerated migratory term to dominates at larger radii. Thus the stellar wind's thickness approaches a saturation when the pressure contribution becomes negligible compared to accelerated migratory term, indicating a transition to migration-dominated stellar wind dynamics. However, in non-accelerated migrating CCSNe case, the stellar wind influenced solely by AGN disk pressure, resulting no significant thickness difference between inner and outer sides.

Accelerated migrating CCSNe more readily generate anisotropic stellar winds on both sides, where the thickness difference between inner and outer sides is significantly larger in the presence of accelerated case than in the non-accelerated case. A central SMBH with a mass of $10^9~M_{\odot}$ more efficiently generates stellar winds with a larger thickness difference between the two sides than the lower massive SMBHs before the stellar winds stabilize, the difference increase progressively and ultimately stabilize. More massive BHs diminish the anisotropy and decrease the total luminosity. Accelerated migrating CCSNe can undergo anisotropic explosion under the influence of AGN disk, whereas no much difference is observed for the non-accelerated cases even taking into account the pressure gradients of the AGN disks, and their anisotropic luminosities are pronounced at large radii. Moreover, higher burst energies, greater ejecta mass, and higher accretion efficiency do not significantly enhance the anisotropy of CCSNe.

Although our study identifies anisotropic explosions in accelerated migrating CCSNe, we focus solely on the effects of differences in stellar winds and approximate the resulting luminosity variations by modeling two separate isotropic bursts within a simplified one-dimensional framework. In future work, we will consider the direction of mass ejection from the progenitor \citep[e.g.,][]{2010ApJ...708.1703M}, the potential presence such as the magnetic fields, and the asymmetric shocks break in AGN disk, where the initially shot out material expands out to hide the subsequently breaking out annuli, quenching the detectable light \citep{2021MNRAS.508.5766I,2021MNRAS.507..156G}. One can conjecture that the newborn compact objects will be popped out from the center of the anisotropic CCSNe, which may be a new type of migrations; and the gas thrown out by explosion might fall back again to form a new star before falling into the SMBHs.

CCSNe occurring in AGN disks may serve as sources of gravitational waves \citep[GWs, e.g.,][]{2016PhRvD..94l3012P,2018ApJ...861...10M}, particularly due to anisotropy in the explosion mechanism and mass ejection. Detecting these GWs alongside their electromagnetic counterparts could provide critical insights into the progenitor's environment and explosion dynamics. Additionally, polarization holds significant importance for studying the radiation mechanisms and the characteristics of the radiation propagation medium. For the geometrically asymmetrical CCSNe in AGN disks, the polarization detection might be an effective method for validating our model.

\acknowledgments
We thanks the anonymous referee for the very helpful suggestions and comments. This work was supported by the National Key R\&D Program of China under grant 2023YFA1607902, the National Natural Science Foundation of China under grants 12173031, 12494572, 12221003, 12321003, 12273113, and 12303049, the Fundamental Research Funds for the Central Universities (No. 20720240152), the Fund of National Key Laboratory of Plasma Physics (No. 6142A04240201), the Youth Innovation Promotion Association (grant No. 2023331), the China Postdoctoral Science Foundation (grant No. 2024M751769), and the China Manned Space Program with grant No. CMS-CSST-2025-A13.

\end{document}